\documentclass[aps,prl,twocolumn,showpacs,superscriptaddress,groupedaddress,reprint]{revtex4}  
\usepackage{graphicx}
\usepackage{dcolumn}
\usepackage{bm}
\usepackage{graphicx}  
\usepackage{dcolumn}   
\usepackage{bm}        
\usepackage{amssymb}   
\usepackage{color}

\newcommand{\COR}[1]{\textcolor{black}{#1}}

\begin{document}

\preprint{AIP/123-QED}

\title[Interaction-Free Ghost Imaging]{Interaction-Free Ghost-Imaging of Structured Objects}

\author{Yingwen \surname{Zhang}}
\affiliation{Physics Department, Centre for Research in Photonics, University of Ottawa, Advanced Research Complex, 25 Templeton, Ottawa ON Canada, K1N 6N5}

\author{Alicia \surname{Sit}}
\affiliation{Physics Department, Centre for Research in Photonics, University of Ottawa, Advanced Research Complex, 25 Templeton, Ottawa ON Canada, K1N 6N5}
	
\author{Fr\'ed\'eric Bouchard}
\affiliation{Physics Department, Centre for Research in Photonics, University of Ottawa, Advanced Research Complex, 25 Templeton, Ottawa ON Canada, K1N 6N5}

\author{Hugo Larocque}
\affiliation{Physics Department, Centre for Research in Photonics, University of Ottawa, Advanced Research Complex, 25 Templeton, Ottawa ON Canada, K1N 6N5}

\author{Eliahu \surname{Cohen}}
\affiliation{Physics Department, Centre for Research in Photonics, University of Ottawa, Advanced Research Complex, 25 Templeton, Ottawa ON Canada, K1N 6N5}
\affiliation{H.H. Wills Physics Laboratory, University of Bristol, Tyndall Avenue, Bristol, BS8 1TL, UK}
\affiliation{Iyar, The Israeli Institute for Advanced Research, POB 651, Zichron, Ya’akov, 3095303, Israel}

\author{Avshalom C. \surname{Elitzur}}
\affiliation{Iyar, The Israeli Institute for Advanced Research, POB 651, Zichron, Ya’akov, 3095303, Israel}
\affiliation{Institute for Quantum Studies, Chapman University, Orange, CA 92866, USA}

\author{James L. Harden}
\affiliation{Physics Department, Centre for Research in Photonics, University of Ottawa, Advanced Research Complex, 25 Templeton, Ottawa ON Canada, K1N 6N5}

\author{Robert W. Boyd}
\affiliation{Physics Department, Centre for Research in Photonics, University of Ottawa, Advanced Research Complex, 25 Templeton, Ottawa ON Canada, K1N 6N5}
\affiliation{Institute of Optics, University of Rochester, Rochester, New York, 14627, USA}

\author{Ebrahim Karimi}
\email{ekarimi@uottawa.ca}
\affiliation{Physics Department, Centre for Research in Photonics, University of Ottawa, Advanced Research Complex, 25 Templeton, Ottawa ON Canada, K1N 6N5}

\date{\today}

\begin{abstract}
Quantum  -- or classically correlated -- light can be employed in various ways to improve resolution and measurement sensitivity. In an ``interaction-free'' measurement, a single photon can be used to reveal the presence of an object placed within one arm of an interferometer without being absorbed by it. This method has previously been applied to imaging. With a technique known as ``ghost imaging'', entangled photon pairs are used for detecting an opaque object with significantly improved signal-to-noise ratio while preventing over-illumination. Here, we integrate these two methods to obtain a new imaging technique which we term ``interaction-free ghost-imaging" that possesses the benefits of both techniques. While maintaining the image \COR{quality} of conventional ghost-imaging, this new technique is also sensitive to phase and polarisation changes in the photons introduced by a structured object. Furthermore, thanks to the ``interaction-free'' nature of this new technique, it is possible to reduce the number of photons required to produce a clear image of the object (which could be otherwise damaged by the photons) making this technique superior for probing light-sensitive materials and biological tissues.
\end{abstract}

\pacs{06.90.+v, 03.67.Mn}
\keywords{Quantum Imaging; Interaction-Free; Ghost Imaging.}
\maketitle


Quantum metrology enables single photons, entangled photon pairs, or multi-photon quantum states to be used for enhancing the resolution of measurements~\cite{Giovannetti:2011aa}. Such states are now applied in several imaging schemes such as interaction-free imaging~\cite{Elitzur1993}, ghost-imaging~\cite{Padgett20160233,Shapiro2012}, and imaging using N00N-states~\cite{Mitchell:2004aa}. Interaction-free imaging involves a single photon going through an interferometer and indicating an object's presence or other physical properties of that object by the {\it absence} of a visible interaction with it~\cite{Elitzur1993}. If undisturbed in the interferometer, the photon interferes with itself and exits through only one output port leaving the other ``dark". If, however, an object is placed in one arm of the interferometer, then the photon's interference is disturbed. One quarter of the time, the detection of the photon in the supposedly dark output port of the interferometer will occur. The key aspect of this technique relies on the fact that photons detected in the dark output port have never interacted with the object, yet can still reveal its presence. Enhancements of the original method~\cite{White1998} involve the quantum Zeno effect~\cite{Kwiat1995,Kwiat1999}, as well as more elaborate schemes based on induced coherence without induced emission~\cite{Lemos2014}. Similar interaction-free measurements have also found many applications in quantum computation and communication~\cite{Hosten2005,Cao2017}, stabilizing ultracold atoms~\cite{Peise2015}, orbital angular momentum spectrometry~\cite{Bierdz2011} and optical switching~\cite{McCusker2013}.

Entangled photon pairs produced through spontaneous parametric down-conversion (SPDC) can be used to produce images with high signal-to-noise ratio (SNR). In ghost imaging~\cite{Klyshko1994,Pittman1995,Aspden2013}, one of the down-converted photons is used to illuminate an object and is captured by a bucket detector. The other is sent through a different path to a camera. By registering only coincidence events between the camera and the bucket detector, an image of the object is formed on the camera even though the photons collected by the camera have never interacted with the object. This also gives images a high SNR and allows images to be obtained with an average of fewer than one detected photon per image pixel~\cite{morris2015}. Ghost imaging can also be achieved with classical light~\cite{Bennink2002,Ferri2005,Bennink2004}. Ghost imaging schemes, relying on non-degenerate down-converted photon pairs, have also been demonstrated~\cite{aspden2015}. A major disadvantage of ghost imaging is its inability to image birefringent and phase-only objects~\cite{Bennink2002}. Another quantum imaging technique making use of entangled photons, the sub-shot-noise imaging protocol, had also been demonstrated to significantly improve SNR in imaging~\cite{Brida2010,Samantaray2017}.

In this Letter, we merge the two ideas and demonstrate interaction-free ghost imaging (IFGI) which possesses the benefits of both techniques. In IFGI, an interferometer is built along the path of the photon used to interrogate the object, \COR{a bucket detector is placed at each of the exit ports. If no object is present, then photons will only be registered in the exit port with constructive interference. Once, however, an object is placed in one arm of the interferometer, the interference is disturbed and both bucket detectors may detect photons. By subtracting the image obtained in the destructive interference port from that of the constructive port, we can obtain an image with the same or even better quality compared to that of conventional ghost imaging (CGI) with less photons interacting with the object.} This feature could be of great importance in imaging objects that display great sensitivity to light, such as certain organic tissues. Furthermore, we show that IFGI is very sensitive to phase and polarisation changes in the beam and can be used to image birefringent and phase-only objects.

\COR{Ideally, for CGI, when $N$ photons are used to probe an object, all $N$ photons will be absorbed/scattered by the object giving a change in photon number at the detector of $\Delta N_{\text{CGI}} = -N$. Now, for IFGI, let us assume a Mach-Zehnder interferometer with the input beam splitter (BS) having reflectivity $R$ and transmissivity $T$, and the exit BS having reflectivity $T$ and transmissivity $R$. If $N$ photons are in the beam when entering the first BS and the object is placed in the transmission arm of the interferometer, then in the constructive exit port, a change in photon number of $\Delta N_C=N(R^2-1)$ will be observed. In the destructive exit port a change of $\Delta N_D= NRT$ will be observed. Subtracting $\Delta N_D$ from $\Delta N_C$ gives}
\begin{equation}
	\COR{\Delta N_{\text{IFGI}} = N(R^2-RT-1)},
	\label{NIFGI}
\end{equation}
\COR{so to have $\Delta N_{\text{IFGI}} = \Delta N_{\text{CGI}}$ will require $R=T=0.5$ (ignoring the trivial case of $T=1$). This means that only $N/2$ photons will have interacted with the object, effectively reducing the interaction by half compared to CGI, while maintaining the same sensitivity. Moreover, when $T>R$, one can obtain $\Delta N_{\text{IFGI}} > \Delta N_{\text{CGI}}$ with a maximal $\Delta N_{\text{IFGI}}=1.125\Delta N_{\text{CGI}}$ (this case can be achieved with $T=0.75$ and $R=0.25$). Therefore, one can achieve an increase in sensitivity by a factor of $1/8$ by using IFGI with a $25\%$ reduction of the photon interaction with the object.}

\begin{figure}
	\centerline{\scalebox{1}{\includegraphics[width=0.5\textwidth]{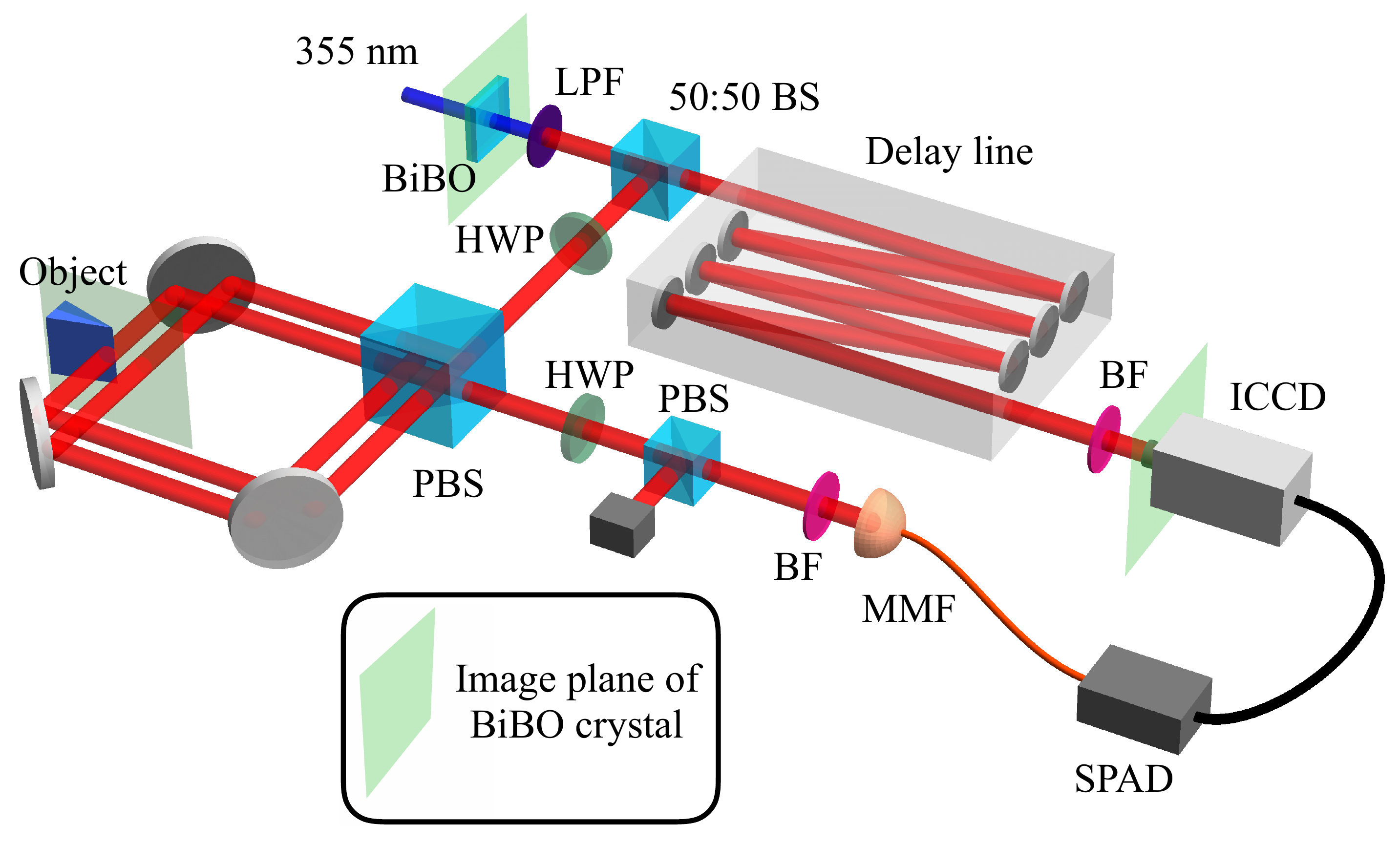}}}
	\caption{Schematic illustration of the experimental setup for IFGI. Entangled photon pairs are generated at a nonlinear crystal (BiBO). One photon is sent into an interferometer with the object to be interrogated; the second photon is sent to a camera through a delay line where the image of the object is formed. Imaging lenses are not shown in the schematic. Figure legends: BiBO - 0.5~mm thick bismuth triborate crystal; LPS - Long-Pass Filter; BS - Beam Splitter; HWP - Half-wave plate; PBS - polarizing beam splitter; BF - Bandpass Filter; MMF - Multi-Mode Fibre; SPAD - Single Photon Avalanche Diode, ICCD - Intensified CCD camera}
	\label{setup}
\end{figure}

Our experimental setup for implementing IFGI is shown in Fig.~\ref{setup}. \COR{A 0.5-mm-thick Type-I bismuth triborate crystal (BiBO) is pumped by a $100$~mW, $355$~nm beam to generate entangled photon pairs via SPDC. The $355$~nm pump beam is afterwards filtered out with a long-pass filter.} The photon pairs are separated by a 50:50 BS where one photon is sent into an interferometer to probe an object and the other to an intensified CCD (ICCD) camera where the image of the probed object is to be viewed. A Sagnac interferometer is constructed to ensure better stability of the interference. The clockwise and anti-clockwise beam paths of the interferometer are slightly displaced from each other such that the object can be placed in just one of the paths. \COR{A variable beam splitter (VBS) -- composed of a half-wave plate (HWP) and a polarizing beam splitter (PBS) -- was used to construct the interferometer in order to give us control over the amount of photons sent to probe the object. Imaging lenses are used to ensure that the ICCD camera and the object are in the same image-plane of the BiBO crystal. A second VBS is placed in the exit port of the interferometer allowing us to observe either constructive or destructive interference. A bucket detector composed of a multi-mode fibre (MMF) with a core diameter of $200$~$\mu$m connected to a single photon avalanche diode (SPAD) is placed behind the VBS.} The ICCD camera is triggered by the SPAD to detect coincidence events between the two arms. To compensate for the timing delay caused by the electronics, an image-preserving delay line of $24$~m length is placed in the path of the photons incident on the ICCD camera. The gating time on the ICCD is set to $5$~ns. Bandpass filters of $710\pm5$~nm are used so that only degenerate photon pairs are detected. \COR{Our IFGI setup can be converted into a CGI setup by simply adjusting the VBS to be fully transmissive. It is important to note that ideally one would like to place a bucket detector at each of the constructive and destructive output ports of the second VBS and have them trigger the same camera in parallel so two images, one for each output port, can be acquired at the same time. However, such a camera is not currently available to us, so we have opted for the current setup as a proof of principle.} 

\COR{To make a fair comparison of the sensitivity and image quality obtained using IFGI and CGI, we use the difference in the average number of photons detected per pixel for a region inside and outside of the object, i.e.,}
\begin{equation}
	\COR{\Delta N = \bar{N}_{\text{out}} -\bar{N}_{\text{in}}}.
\end{equation}
\COR{Here, we did not employ the more commonly used quantity, namely, the contrast or visibility, defined as $V=(\bar{N}_{\text{out}} -\bar{N}_{\text{in}})/(\bar{N}_{\text{out}}+\bar{N}_{\text{in}})$ to compare the images. As previously discussed, making a fair comparison between IFGI and CGI would require a direct comparison between the number of photons used to create the image. However, $V$ is a normalized quantity and can give the same value provided that its ratio is maintained irrespective of the number of photons, and therefore cannot be used as a figure of merit to compare the two methods.}

When making coincidence measurements, one also has to consider the subtraction of accidental events caused by background light, detector noise, etc. To account for accidental events during CGI measurements, we adjust the delay of our ICCD such that, when triggered by the bucket detector, it opens its shutter outside the coincidence window thereby registering only accidental events. The image obtained this way is subtracted off as background. In IFGI measurements, we have an additional background contribution due to imperfect interference. This is caused by a combination of an imperfect beam splitter and interferometer alignment. To account for this, we simply used the image obtained in the destructive port when there is no object in the interferometer as the background.

\begin{figure}[h]
	\centerline{\scalebox{1}{\includegraphics[width=0.5\textwidth]{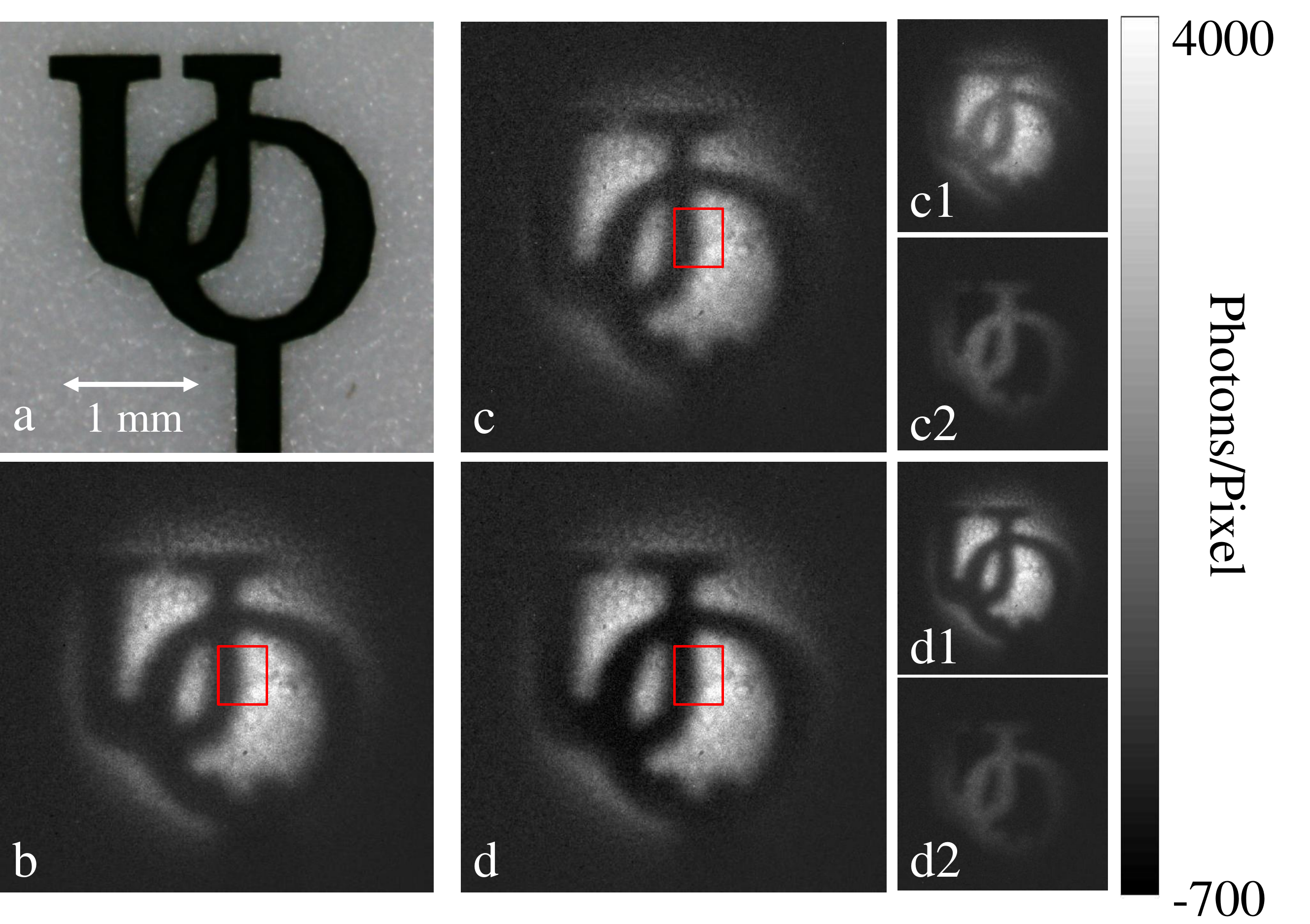}}}
	\caption{\COR{A laser cut metal signs of the letters UO - University of Ottawa, with $\sim$2~mm diameter (a), was imaged using CGI (b) and IFGI with the VBS adjusted to have 50:50 (c) and 25:75 (d) (R:T) ratio. (c1)(d1) and (c2)(d2) are the images obtained from the constructive and destructive ports of the interferometer respectively. (c) is obtained by subtracting (c2) from (c1) and similarly (d) is obtained by subtracting (d2) from (d1). $\Delta N$ of the image is determined in the region enclosed by the red square.  $\Delta N$ of the CGI image (b) is determined to be 2090~photons/pixel and for IFGI is 2045 and 2487~photons/pixel for the 50:50 (c) and 25:75 (d) cases respectively. The integration time of the images was 5 minutes.} }
	\label{UO}
\end{figure}

In Fig.~\ref{UO} we show a comparison of images obtained using CGI and IFGI. \COR{The imaged object is a $\sim$2-mm-diameter laser cut metal sign of the letters ``UO" (Fig.~\ref{UO}-(a)). An image of the sign is taken with CGI as shown in Fig.~\ref{UO}-(b), and a $\Delta N$ of 2090~photons/pixel was obtained. In Fig.~\ref{UO}-(c1) and (c2), we see the images taken in the constructive and destructive interference port, respectively, with the VBS set to 50:50, and Fig.~\ref{UO}-(c) is the image obtained from (c1) subtracting (c2). For Fig.~\ref{UO}-(c), we obtained a $\Delta N$ of 2045~photons/pixel, a number that is comparable to that obtained from CGI. Figure~\ref{UO}-(d) is generated via the subtraction of Fig.~\ref{UO}-(d2) from (d1), the constructive and destructive interference port, respectively, when the VBS is set to 25:75 (R:T). A $\Delta N$ of 2487~photons/pixel was obtained for Fig.~\ref{UO}-(d), agreeing with the predicted $1/8$ increase over CGI.} Potentially, one can further reduce the amount of absorbed photons to zero by implementing the quantum Zeno effect~\cite{Kwiat1995,Kwiat1999}.
	
\COR{A minor technical drawback of IFGI seen in these results is that it will have more background noise compared to CGI, caused by the combined noise when the images are being added/subtracted. To create the final image using IFGI, three different images have to be added/subtracted from each other, namely the images from the constructive and destructive ports of the VBS and the background due to the imperfect interference. This would theoretically increase the background noise by a factor of $\sqrt{3}$. The standard deviation in the background events determined for CGI in Fig.~\ref{UO}-(b) was 64~photons/pixel and that for IFGI in Figs.~\ref{UO}-(c) and (d) are 116 and 104~photons/pixel respectively, which agrees with the predicted noise of $\sqrt{3}\times64=111$~photons/pixel.}

\begin{figure}[h]
	\centerline{\scalebox{1}{\includegraphics[width=0.5\textwidth]{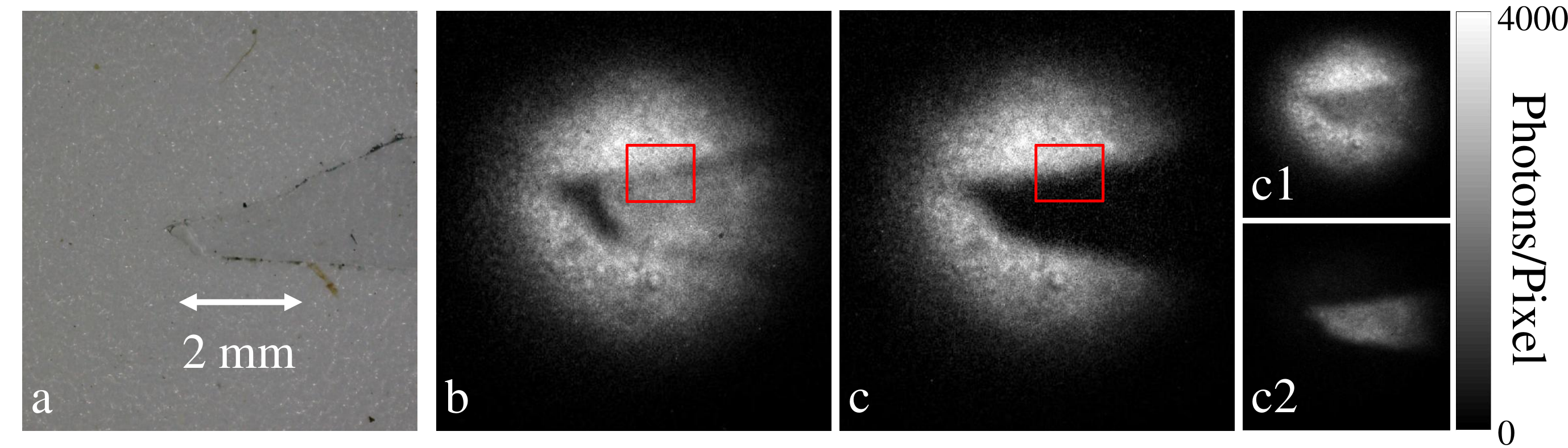}}}
	\caption{\COR{A 0.15-mm-thick glass shard (a) is imaged with CGI (b) and IFGI (c). (c) is obtained from subtracting the image in the destructive port (c2) of the interferometer from that of the constructive port (c1). $\Delta N$ for the images of the shard is determined for the region enclosed by the red square giving 993~photons/pixel for (b) and 2391~photons/pixel for (c). The integration time was 5 minutes.}}
	\label{transparent}
\end{figure}

Another advantage of IFGI when compared to CGI is in its sensitivity to changes in phase and polarisation of the photons. In Fig.~\ref{transparent}, a $\sim$0.15-mm-thick glass shard (Fig.~\ref{transparent}-(a)) is placed in the beam and imaged by CGI (Fig.~\ref{transparent}-(b)) and IFGI (Fig.~\ref{transparent}-(c)). Using CGI, we can see the area blocked by the glass shard is dimmer. This would be caused by the back reflection and scattering of photons in the glass. If the glass shard is anti-reflection treated, then it should be almost totally invisible to CGI except for perhaps the edges. With IFGI, the glass shard is clearly visible, because the phase shift introduced by the glass disturbs the interference. \COR{We have also determined $\Delta N$ of the images, obtaining 993~photons/pixel for CGI and 2391~photons/pixel for IFGI. Clearly a much larger $\Delta N$ for IFGI.}

\begin{figure}[h]
	\centerline{\scalebox{1}{\includegraphics[width=0.5\textwidth]{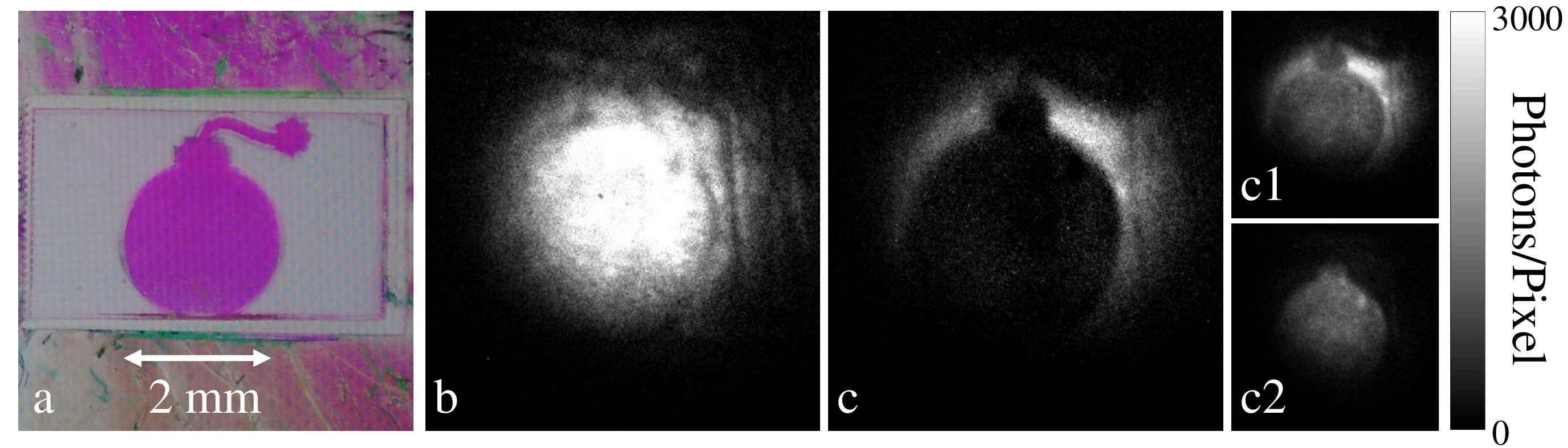}}}
	\caption{\COR{A ``bomb" pattern imprinted on a liquid crystal device (a) imaged with CGI (b) and IFGI (c). (c) is obtained from subtracting the image in the destructive port (c2) of the interferometer from that of the constructive port (c1). The integration time was 5 minutes.}}
	\label{Bomb}
\end{figure}

The sensitivity to polarisation of IFGI is demonstrated in Fig.~\ref{Bomb}. Here, a pattern of a ``bomb" is imprinted on a liquid crystal device \cite{Larocque2016}; an image of the device placed between two cross polarisers is shown in Fig.~\ref{Bomb}-(a). Any linearly polarised photons passing through the pattern will have their polarisation rotated by 90$^\circ$. \COR{For a CGI setup, there are normally no polarisation sensitive elements so we had to remove both VBSs from the setup when the CGI image was taken. The CGI image taken without any VBS is shown in Fig.~\ref{Bomb}-(b) and this polarisation change is clearly not observed. However, the bomb pattern is clearly visible in Fig.~\ref{Bomb}-(c) when IFGI is used, indicating a sensitivity to polarisation change.  We have not compared $\Delta N$ between the two methods here as the bomb is clearly invisible to CGI. Polarisation sensitivity can certainly be added to CGI simply by placing a polariser after the object, but what we want to demonstrate here is that polarisation sensitivity is an intrinsic feature to IFGI.} To compensate for the phase change introduced by the liquid crystal device, a glass plate of similar thickness is inserted in the other interferometer arm and with its insertion angle adjusted such that a good interference is observed by the bucket detector. 

On the fundamental side, our demonstration proves once again the robustness of quantum nonlocality and interaction-free measurement, even when combined together. Although ghost-imaging could similarly work for classically correlated light (with a predetermined direction of the photons)~\cite{Bennink2002,Ferri2005,Bennink2004}, we have used here nonlocally correlated photons (with the aim of further utilizing in the future the quantum features of the photons, e.g. when employing N00N or squeezed states). We have therefore found evidence that interaction-free measurement performed with the aid of one photon leads to ``collapse", thereby affecting its remote entangled partner. The nonlocal features are clearly apparent in previous proposals along these lines~\cite{Jack2009,Angelo2005,Elitzur2001,Elitzur2014}, but in this experiment we chose to focus on the interaction-free aspects. The introduction of interaction-free imaging makes the present experiment genuinely quantum~\cite{Robens2017}. Moreover, it is ``doubly-nonlocal'' since it employs both the familiar quantum nonlocality made possible through entanglement, as well as the more subtle nonlocality of two wavepackets corresponding to the same particle \cite{Hardy1994}. Notice that the latter enables to sense not only the presence of an object, but also its properties, without necessarily changing it (by virtue of a quantum counterfactual). Eventually, the unwarranted effect of the photon on the probed object, e.g. a delicate biological tissue, can be largely avoided, yet the state of the photon {\it does} change. This curious phenomenon, namely one party affecting the other without a reciprocal action, was studied earlier on the basis of an underlying mechanism termed ``quantum oblivion'' \cite{Elitzur2014}. That analysis turned out to underlie some other related phenomena too. Indeed, when a photon is employed in interaction-free measurement, its relative phase changes, but the ``bomb'' system, apparently oblivious to the interaction, does not change. It is this non-reciprocity that enables gathering information while inducing reduced disturbance.  

In summary, we have demonstrated a new imaging method termed Interaction-Free Ghost-Imaging (IFGI) and have shown how it can serve as a useful ghost imaging technique to observe light sensitive structured objects. When compared to CGI, \COR{a similar and even improved image quality can be obtained with IFGI while reducing the number of photons probing the object.} The extra interferometer in the IFGI setup also makes the proposed technique sensitive to phase shifts and polarisation changes of light introduced by the observed object, whose effects are mostly invisible to CGI. IFGI would prove useful in imaging biological samples where changes in the sample thickness and composition can be observed. In order to view smaller samples, such as biological cells, one can focus the pump beam onto the non-linear crystal giving the SPDC beam a much smaller profile. However, this reduces the Rayleigh range of the SPDC photons, making the image plane of the object much more difficult to find on the camera. This sensitivity to phase and polarisation can certainly also be achieved using ghost-imaging with classical light~\cite{Zhang2014}. However, the ``interaction-free'' nature of this experiment is a purely quantum phenomenon and can be used to reduce the number of photons required to probe a light sensitive object which is otherwise not achievable with classical light. In light of the above demonstrative results, additional promising avenues are waiting to be explored, such as extending the technique to X-rays \cite{Schori2017}, where reducing the amount of absorbed radiation by the tissues is vital.

This work was supported by the Canada Research Chairs (CRC) and Canada Excellence Research Chairs (CERC) Program. E.C. was supported by ERC AdG NLST. The authors acknowledge Robert Fickler, Peter Morris, Miles Padgett and John Sipe for helpful discussions \COR{and would like to thank the referees for their helpful comments}.

\bibliographystyle{unsrt}
\bibliography{references}

\begin{thebibliography}{10}

\bibitem{Giovannetti:2011aa}
Vittorio Giovannetti, Seth Lloyd, and Lorenzo Maccone.
\newblock Advances in quantum metrology.
\newblock {\em Nat Photon}, 5(4):222--229, 04 2011.

\bibitem{Elitzur1993}
Avshalom~C. Elitzur and Lev Vaidman.
\newblock Quantum mechanical interaction-free measurements.
\newblock {\em Foundations of Physics}, 23(7):987--997, Jul 1993.

\bibitem{Padgett20160233}
Miles~J. Padgett and Robert~W. Boyd.
\newblock An introduction to ghost imaging: quantum and classical.
\newblock {\em Philosophical Transactions of the Royal Society of London A:
  Mathematical, Physical and Engineering Sciences}, 375(2099), 2017.

\bibitem{Shapiro2012}
Jeffrey~H. Shapiro and Robert~W. Boyd.
\newblock The physics of ghost imaging.
\newblock {\em Quantum Information Processing}, 11(4):949--993, Aug 2012.

\bibitem{Mitchell:2004aa}
M.~W. Mitchell, J.~S. Lundeen, and A.~M. Steinberg.
\newblock Super-resolving phase measurements with a multiphoton entangled
  state.
\newblock {\em Nature}, 429:161 EP --, 05 2004.

\bibitem{White1998}
Andrew~G. White, Jay~R. Mitchell, Olaf Nairz, and Paul~G. Kwiat.
\newblock ``interaction-free'' imaging.
\newblock {\em Phys. Rev. A}, 58:605--613, Jul 1998.

\bibitem{Kwiat1995}
Paul Kwiat, Harald Weinfurter, Thomas Herzog, Anton Zeilinger, and Mark~A.
  Kasevich.
\newblock Interaction-free measurement.
\newblock {\em Phys. Rev. Lett.}, 74:4763--4766, Jun 1995.

\bibitem{Kwiat1999}
P.~G. Kwiat, A.~G. White, J.~R. Mitchell, O.~Nairz, G.~Weihs, H.~Weinfurter,
  and A.~Zeilinger.
\newblock High-efficiency quantum interrogation measurements via the quantum
  zeno effect.
\newblock {\em Phys. Rev. Lett.}, 83:4725--4728, Dec 1999.

\bibitem{Lemos2014}
Gabriela~Barreto Lemos, Victoria Borish, Garrett~D Cole, Sven Ramelow, Radek
  Lapkiewicz, and Anton Zeilinger.
\newblock Quantum imaging with undetected photons.
\newblock {\em Nature}, 512(7515):409--412, 2014.

\bibitem{Hosten2005}
Onur Hosten, Matthew~T. Rakher, Julio~T. Barreiro, Nicholas~A. Peters, and
  Paul~G. Kwiat.
\newblock Counterfactual quantum computation through quantum interrogationt.
\newblock {\em Nature}, 439:949--952, 2005.

\bibitem{Cao2017}
Yuan Cao, Yu-Huai Li, Zhu Cao, Juan Yin, Yu-Ao Chen, Hua-Lei Yin, Teng-Yun
  Chen, Xiongfeng Ma, Cheng-Zhi Peng, and Jian-Wei Pan.
\newblock Direct counterfactual communication via quantum zeno effect.
\newblock {\em Proceedings of the National Academy of Sciences},
  114(19):4920--4924, 2017.

\bibitem{Peise2015}
J.~Peise, B.~L{\"u}cke, L.~Pezz{\'e}, F.~Deuretzbacher, W.~Ertmer, J.~Arlt,
  A.~Smerzi, L.~Santos, and C.~Klempt.
\newblock Interaction-free measurements by quantum zeno stabilization of
  ultracold atoms.
\newblock {\em Nature communications}, 6:6811, 2015.

\bibitem{Bierdz2011}
Paul Bierdz and Hui Deng.
\newblock A compact orbital angular momentum spectrometer using quantum zeno
  interrogation.
\newblock {\em Opt. Express}, 19(12):11615--11622, Jun 2011.

\bibitem{McCusker2013}
Kevin~T. McCusker, Yu-Ping Huang, Abijith~S. Kowligy, and Prem Kumar.
\newblock Experimental demonstration of interaction-free all-optical switching
  via the quantum zeno effect.
\newblock {\em Phys. Rev. Lett.}, 110:240403, Jun 2013.

\bibitem{Klyshko1994}
A.~V. Belinskii and D.~N. Klyshko.
\newblock Two-photon optics: diffractlon, holography, and transformation of
  two-dimensional signals.
\newblock {\em JETP}, 78:259, 1994.

\bibitem{Pittman1995}
T.~B. Pittman, Y.~H. Shih, D.~V. Strekalov, and A.~V. Sergienko.
\newblock Optical imaging by means of two-photon quantum entanglement.
\newblock {\em Phys. Rev. A}, 52:R3429--R3432, Nov 1995.

\bibitem{Aspden2013}
Reuben~S Aspden, Daniel~S Tasca, Robert~W Boyd, and Miles~J Padgett.
\newblock Epr-based ghost imaging using a single-photon-sensitive camera.
\newblock {\em New Journal of Physics}, 15(7):073032, 2013.

\bibitem{morris2015}
Peter~A. Morris, Reuben~S. Aspden, Jessica E.~C. Bell, Robert~W. Boyd, and
  Miles~J. Padgett.
\newblock Imaging with a small number of photons.
\newblock {\em Nature Communications}, 6:5913, May 2015.

\bibitem{Bennink2002}
Ryan~S. Bennink, Sean~J. Bentley, and Robert~W. Boyd.
\newblock ``two-photon'' coincidence imaging with a classical source.
\newblock {\em Phys. Rev. Lett.}, 89:113601, Aug 2002.

\bibitem{Ferri2005}
F.~Ferri, D.~Magatti, A.~Gatti, M.~Bache, E.~Brambilla, and L.~A. Lugiato.
\newblock High-resolution ghost image and ghost diffraction experiments with
  thermal light.
\newblock {\em Phys. Rev. Lett.}, 94:183602, May 2005.

\bibitem{Bennink2004}
Ryan~S. Bennink, Sean~J. Bentley, Robert~W. Boyd, and John~C. Howell.
\newblock Quantum and classical coincidence imaging.
\newblock {\em Phys. Rev. Lett.}, 92:033601, Jan 2004.

\bibitem{aspden2015}
Reuben~S Aspden, Nathan~R Gemmell, Peter~A Morris, Daniel~S Tasca, Lena
  Mertens, Michael~G Tanner, Robert~A Kirkwood, Alessandro Ruggeri, Alberto
  Tosi, Robert~W Boyd, et~al.
\newblock Photon-sparse microscopy: visible light imaging using infrared
  illumination.
\newblock {\em Optica}, 2(12):1049--1052, 2015.

\bibitem{Brida2010}
G.~Brida, M.~Genovese, and I.~Ruo~Berchera.
\newblock Experimental realization of sub-shot-noise quantum imaging.
\newblock {\em Nature Photonics}, 4:227, 2010.

\bibitem{Samantaray2017}
Nigam Samantaray, Ivano Ruo-Berchera, Alice Meda, and Marco Genovese.
\newblock Realization of the first sub-shot-noise wide field micscope.
\newblock {\em Light: Science \& Applications}, 6:e17005, 2017.

\bibitem{Larocque2016}
Hugo Larocque, J\'{e}r\'{e}mie Gagnon-Bischoff, Fr\'{e}d\'{e}ric Bouchard,
  Robert Fickler, Jeremy Upham, Robert~W Boyd, and Ebrahim Karimi.
\newblock Arbitrary optical wavefront shaping via spin-to-orbit coupling.
\newblock {\em Journal of Optics}, 18(12):124002, 2016.

\bibitem{Jack2009}
B.~Jack, J.~Leach, J.~Romero, S.~Franke-Arnold, M.~Ritsch-Marte, S.~M. Barnett,
  and M.~J. Padgett.
\newblock Holographic ghost imaging and the violation of a bell inequality.
\newblock {\em Phys. Rev. Lett.}, 103:083602, Aug 2009.

\bibitem{Angelo2005}
Milena D'Angelo, Alejandra Valencia, Morton~H. Rubin, and Yanhua Shih.
\newblock Resolution of quantum and classical ghost imaging.
\newblock {\em Phys. Rev. A}, 72:013810, Jul 2005.

\bibitem{Elitzur2001}
Avshalom~C. Elitzur and Shahar Dolev.
\newblock Nonlocal effects of partial measurements and quantum erasure.
\newblock {\em Phys. Rev. A}, 63:062109, May 2001.

\bibitem{Elitzur2014}
Avshalom~C. Elitzur and Eliahu Cohen.
\newblock Quantum oblivion: A master key for many quantum riddles.
\newblock {\em International Journal of Quantum Information},
  12(07n08):1560024, 2014.

\bibitem{Robens2017}
Carsten Robens, Wolfgang Alt, Clive Emary, Dieter Meschede, and Andrea Alberti.
\newblock Atomic ``bomb testing'': the elitzur--vaidman experiment violates the
  leggett--garg inequality.
\newblock {\em Applied Physics B}, 123:12, Dec 2016.

\bibitem{Hardy1994}
Lucien Hardy.
\newblock Nonlocality of a single photon revisited.
\newblock {\em Phys. Rev. Lett.}, 73:2279--2283, Oct 1994.

\bibitem{Zhang2014}
De-Jian Zhang, Qiang Tang, Teng-Fei Wu, Hao-Chuan Qiu, De-Qin Xu, Hong-Guo Li,
  Hai-Bo Wang, Jun Xiong, and Kaige Wang.
\newblock Lensless ghost imaging of a phase object with pseudo-thermal light.
\newblock {\em Applied Physics Letters}, 104(12):121113, 2014.

\bibitem{Schori2017}
A.~Schori and S.~Shwartz.
\newblock X-ray ghost imaging with a laboratory source.
\newblock {\em Opt. Express}, 25(13):14822--14828, Jun 2017.

\end{thebibliography}

\end{document}